# MEBS: Multi-task End-to-end Bid Shading for Multi-slot Display Advertising

Zhen Gong[1], Lvyin Niu[2*], Yang Zhao[2*], Miao Xu[2], Zhenzhe Zheng[1], Haoqi Zhang[1], Zhilin Zhang[2], Rongquan Bai[2], Fan Wu[1], Chuan Yu[2], Jian Xu[2], Bo Zheng[2]
Shanghai Jiao Tong University[1], Alibaba Group[2]
{gongzhen, zhengzhenzhe, zhanghaoqi39}@sjtu.edu.cn, fwu@cs.sjtu.edu.cn
{lvyin.nly, jinfeng.zy, xumiao.xm, zhangzhilin.pt, rongquan.br, yuchuan.yc, xiyu.xj, bozheng}@alibaba-inc.com

## Abstract

Online bidding and auction are crucial aspects of the online advertising industry. Conventionally, there is only one slot for ad display and most current studies focus on it. Nowadays, multi-slot display advertising is gradually becoming popular where many ads could be displayed in a list and shown as a whole to users. However, multi-slot display advertising leads to different cost-effectiveness. Advertisers have the incentive to adjust bid prices so as to win the most economical ad positions. In this study, we introduce bid shading into multi-slot display advertising for bid price adjustment with a **M**ulti-task **E**nd-to-end **B**id **S**hading (MEBS) method. We prove the optimality of our method theoretically and examine its performance experimentally. Through extensive offline and online experiments, we demonstrate the effectiveness and efficiency of our method, and we obtain a 7.01% lift in Gross Merchandise Volume, a 7.42% lift in Return on Investment, and a 3.26% lift in ad buy count.

## 1 Introduction

Online advertising serves as one of the most important income sources of Internet companies [8]. The Real-Time Bidding (RTB) paradigm enables the automated matching of advertisements with target audiences and facilitates transactions on a per-impression basis [32]. The RTB system comprises critical components, including advertisers, demand-side platforms (DSP), Ad exchanges (AdX), supply-side platforms (SSP), publishers, and data management platforms (DMP) [36, 37]. The SSP aids publishers in managing and pricing their ad inventories [35]. The DSP assists advertisers by providing bids and targeting potential audiences in the market. The AdX arranges auctions with buyers in DSP and sellers in SSP by allocating ad impression opportunities. Advertisers try to submit optimal bid prices in auctions with the best bidding strategy, which attracts attention from both the industry and academia [1, 5, 40].

Typical display advertising goes with only one ad slot sold each time, which is single-slot display advertising. But there is another emerging scene that many ad slots are shown together and sold together in one auction, which is called multi-slot display advertising (MSDA). The single-slot and multi-slot display advertising are shown in Figure 1. Under the prevalent generalized second-price (GSP) auction setting, each advertiser pays the next highest advertiser's bid [8], and the advertiser with the highest bid wins the first ad position and pays the second-highest bid price in MSDA, the advertiser with the second-highest bid wins the second ad position and pays the third-highest bid price, and so on. Ads in different positions enjoy different levels of user attention. The top positions are usually more attractive [13] but more expensive. And the Click-Through Rate (CTR) varies according to the ad position [6]. In pay-per-click (PPC) mode [8, 30], the ad platform calculates payment according to whether the ad is clicked. However, conversion rates (CVR) do not vary much with ad position [2]. It is easily explicable that CVR is independent of the ad position in the previous displaying page because the conversion happens when the user browses the next page after a click. This leads to different cost–effectiveness for different ad slots. A lower position with a lower price but similar CVR could be a bargain for advertisers, which means advertisers may profit more if competing for those lower cheaper ad slots. Consequently, advertisers are incentivized to cut bid prices and bid on lower but more economical positions.

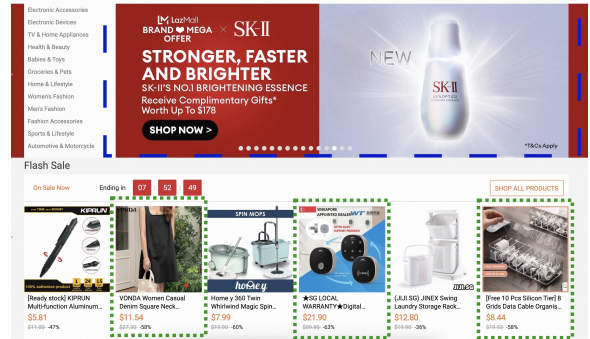

**Figure 1: Examples of single-slot and multi-slot display advertising are shown in the blue dashed and green dotted boxes.**

Bid shading methods are widely adopted to adjust the bid price to avoid overpaying in first-price auctions (FPA). Bid shading estimates shading ratio (the ratio of cutting bid price) or searches for the optimal bid price. It is also potentially feasible to find the most economical bid price with the help of bid shading in multi-slot display advertising, though bid shading in MSDA is rarely studied by the community. Like FPA with bid shading [14], MSDA with bid shading could also be decoupled into three parts: impression valuation, campaign control, and bid shading. Impression valuation estimates the effective cost per mille (eCPM) as ad value, and the eCPM estimation includes the modeling of CTR [25, 41, 42] and CVR [16, 17, 20, 33]. Campaign control generates control signals to adjust ad value according to constraints like budget, and the adjusted ad value serves as the original bid price before bid shading under single-slot GSP setting [11, 12, 44]. Bid shading estimates the optimal bid price to save costs for MSDA. Since impression valuation and campaign control have been widely investigated in previous literature, we focus on bid shading in this work.

There are a few challenges in MSDA involved by bid shading. The optimality of bidding strategy with bid shading in MSDA has not been proven by any previous work. Bid shading lowers down the bid price, and it is only meaningful to the winnable samples whose bid prices are high enough to win auctions. Bid shading also leads to varying CTR. Modeling of winnable samples and CTR change faces the data sparsity problem since winnable samples are rare when bidding is fierce and clicks are far fewer than bidding actions. Optimal bid price has no ground truth since CTR changing is unknowable solely based on dataset and the calibrated predicted CTR (pCTR) is only available after being estimated with a model.

Traditional bid shading methods [9, 43] focus on modeling the optimal bid price distribution, and they estimate the optimal shading ratio or search for the optimal bid price according to the distribution. And they regard the highest competing bid price (the minimum winning price) as the label of optimal bid price. Under single-slot FPA settings, lowering down original bid price to the minimum winning price saves more cost, but the auction is still winnable. All samples in one auction share the same optimal bid price, so the distribution modeling is relatively easy and feasible. But these methods no longer suit MSDA due to ad position effects. Bid shading leads to the change of slots, causing variations in CTR. Consequently, the optimal bid price is no longer solely determined by the minimum winning price. Modeling changes in CTR for bid shading becomes necessary to find the optimal bid price in MSDA.

In this work, we apply a multi-task end-to-end bid shading method to MSDA. We propose an optimal bid shading strategy for MSDA and provide a proof of its optimality. After modeling the change of win rate and CTR during adjusting bid price, MEBS estimates the optimal shading ratio. Multi-task learning helps solve data sparsity problem for pCTR calibration and shading ratio estimation. And we directly maximize the cost saved for advertisers (i.e., surplus) in an end-to-end paradigm, making shading ratio estimation feasible even if the label of optimal shading ratio is unknowable. And we successfully help improve the bidding in MSDA and get promising results from both offline and online experiments.

## 2 Related Work

### 2.1 Bid Shading Methods

There are mainly three kinds of traditional bid shading methods:

Shading Ratio Regression Method [9]: Gligorijevic et al. [9] regard shading ratio estimation as a regression task and directly predict the optimal shading ratio, which is defined as the ratio of optimal bid price label and original bid price. And the optimal shading ratio is utilized as the training label for the regression task. However, this method suffers from data sparsity problem, since the label is only valid for winnable samples [43]. And the label of optimal bid price that the method relies on is unknowable in MSDA.

Two-step Bid Shading Methods [23, 28, 43]: Two-step bid shading has two steps: estimating the winning price distribution, and then searching for the optimal bid price to maximize surplus. The searching process harms inference efficiency and also limits effectiveness in turn. For example, Zhou et al. [43] choose the inferior FwFM [22] model as online model due to online latency constraints, though DeepFM [10] outperforms FwFM in their offline experiments.

Non-parametric Method: MEOW [39] proposes a non-parametric method for bid shading using dynamic binning. It constructs bins to divide impressions into small groups and determines the bid price based on the located bins. This method is space-efficient because only a few bins are required. But it also relies on the optimal bid price label for surplus calculation in its binning process.

### 2.2 Multi-task Learning

ESMM [20] is proposed to relieve the data sparsity problem and attracts a lot of attention. ESMM combines CTR modeling task with CVR estimation task by sharing embedding. The CTR task is trained in the entire space with samples of all impressions, which helps the CVR task with the data sparsity problem. And $ESM^2$ [33] introduces more active post-click user behaviors to further solve data sparsity problem. Zhang et al. [38] and Wang et al. [31] improve multi-task learning from a causal perspective. Apart from above methods, MMoE [19], DUPN [21], and PLE [29] are also widely adopted.

### 2.3 Multi-slot Adverting

Most studies about multi-slot advertising focus on bidding and auction of sponsored search [1, 7, 27]. Auctions for sponsored search are usually held for each individual ad slot instead of all slots on a whole page. So bidding for multi-slot sponsored search does not take the predicted CTR change into account. These keyword-level bidding strategies neglect critical individual impression features (e.g., clicks and positions), which are indispensable in display advertising.

## 3 Preliminaries

We define the necessary notations shown in Table 1. In impression valuation, we estimate the predicted CTR $pCTR(x)$ with features $x$ regardless of the ad slot and obtain ad value $V$. The adjusted ad value $\mu_0^* V$ serves as the unshaded bid price and it is got by campaign control with the optimal control signal $\mu_0^*$. As for bid shading, we generate a bid price shaded by multiplying the shading ratio $r$ with the adjusted ad value, and we know the saved cost called surplus $S$. Bidding consumes the budget $B$ set by advertisers, and its cost is denoted as $C(b)$. The minimum bid price to win the auction is defined as $wp$, which is the bid price that wins the last slot. The total amount of all samples is denoted with $N$, and $N_+$ refers to the amount of winnable samples whose bid prices are higher than $wp$.

Table 1: Notations used throughout the paper.

| Notation | Definition |
| --- | --- |
| $V$ | ad value if clicked (truthful bid price) |
| $\mathbb{E}(V)$ | the expected gained value of an ad |
| $\mu_0^*$ | the optimal control signal |
| $b$ | bid price |
| $S$ | surplus (the cost saved for advertisers) |
| $\mathbb{E}(S)$ | expected surplus |
| $r$ | shading ratio |
| $wp$ | minimum winning price |
| $N$ | total amount of samples |
| $N_+$ | total amount of positive samples |
| $x$ | input features except for bid price |
| $B$ | budget of an advertising campaign |
| $C(b)$ | cost of the bid to win an ad with bid price $b$ |
| $\mathbb{E}(C)$ | the expected cost |
| $P(x, b)$ | win rate (the probability that $b \geq wp$) |
| $y_{\text{WR}}$ | the label of win rate estimation |
| $u_k$ | average CTR in the k-th slot |
| $pCTR(x)$ | pCTR irrespective of ad position |
| $pCTR_k(x, b)$ | pCTR of the k-th slot won with bid price $b$. |
| $y_{\text{Calib}}$ | the label of pCTR calibration |
| $F(x, b)$ | probability of auction is won and ad is clicked |

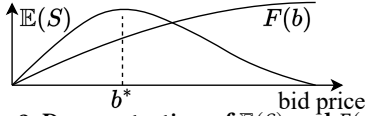

Figure 2: Demonstration of $\mathbb{E}(S)$ and $F(x, b)$.

Apart from shading ratio estimation, MEBS also involves win rate estimation and pCTR calibration. The probability of winning the auction is denoted as the win rate $P(x, b)$. And the label of win rate estimation $y_{\text{WR}}$ indicates whether the auction is won. Following Aggarwal et al. [3], we decompose the pCTR of the ad in the k-th slot into two parts: the ad-specific slot-independent pCTR ($pCTR(x)$) and the slot-specific factor $\frac{u_k}{u_i}$. And $u_k$ is the average CTR in the k-th slot. Assume that we win the k-th slot instead of the original i-th slot after bid shading, the calibrated pCTR for the k-th slot is derived as

$$pCTR_k(x, b) = \frac{u_k}{u_i} pCTR(x). \quad (1)$$

And $y_{\text{Calib}}$ stands for whether the ad is clicked after winning the k-th slot in the auction with bid price $b$. Combining win rate with the calibrated pCTR, we define the probability that the auction is won and the ad is also clicked as $F(x, b) = P(x, b) \cdot pCTR_k(x, b)$. And the expected surplus is

$$\mathbb{E}(S) = (\mu_0^* V - C(b)) F(x, b).$$

The more the bid price is shaded, the more costs are saved, but the less likely to win the auction, and the slot's CTR is lower.

The expected surplus shown in Figure 2 is usually unimodal based on our statistics, and it increases first and then decreases with a rising bid price. The win rate $P(x, b)$ is a positive function, and it is strictly monotonically increasing with the bid price $b$. And the slot-specific pCTR $pCTR_k(x, b)$ is positive and not decreasing with the bid price $b$. So, $F(x, b)$ is also a strictly monotonically increasing and positive function. Besides, cost $C(b)$ is also obviously positive and not monotonically decreasing with bid price $b$.

## 4 Optimal Bidding Strategy

As for MSDA, the bidding objective is to maximize the cumulative expected gained value subject to the finite ad campaign budget. We may formulate the optimization problem as:

$$\max_{b_i} \sum_{i=1}^{N} \mathbb{E}(V_i), \ s.t. \ \sum_{i=1}^{N} \mathbb{E}(C_i) \leq B.$$

The expected gained value can be represented as the multiplication of the probability that the auction is won and the ad is clicked with the ad value, which is $\mathbb{E}(V) = F(x, b)V$. Under PPC mode, the expected cost is $\mathbb{E}(C) = F(x, b)C(b)$ since payments are counted according to clicks. The optimization problem is reformulated as:

$$\max_{b_i} \sum_{i=1}^{N} F(x, b_i) V_i, \ s.t. \ \sum_{i=1}^{N} F(x, b_i) C(b_i) \leq B. \quad (2)$$

The bidding optimization problem in Eq. (2) could be transferred to the following two optimization problems (campaign control and bid shading) shown in Theorem 4.1.

THEOREM 4.1. *The optimal bid price $b^*$ got by bid shading in Eq. (2) satisfies the following two optimal strategies ($\mu_0^*, b^*$) for all sample i of the non-cooperative, non-zero sum, infinite game:*

$$\mu_0^* = \arg\min_{\mu_0} \left( \sum_{i=1}^{N} F(x, b_i) C(b_i) - B \right),$$
$$b_i^* = \arg\max_{b_i} (\mu_0^* V_i - C(b_i)) F(x, b_i), \quad (3)$$

*where $\mu_0^*$ is the optimal control signal for budget pacing in campaign control and $b_i^*$ is the optimal bid price maximizing the expected surplus in bid shading.*

*If the expected surplus $\mathbb{E}(S) = (\mu_0 V_i - C(b_i)) F(x, b_i)$ is a unimodal function with a unique maximum for all samples i and all control signals $\mu_0$, then the bid price $b^*$ for any sample i is the unique optimal bidding strategy solving the optimization problem in Eq. (2).*

PROOF. First, we transform the optimization problem in Eq. (2) to a dual problem. Let $\lambda$ denote the dual variable of budget $B$. The Lagrangian function of optimization problem in Eq. (2) is:

$$\mathcal{L} = \sum_{i=1}^{N} F(x, b_i) V_i - \lambda \left( \sum_{i=1}^{N} F(x, b_i) C(b_i) - B \right)$$
$$= \sum_{i=1}^{N} (V_i - \lambda C(b_i)) F(x, b_i) + \lambda B.$$

According to Lagrangian optimization theory [26], if there exist bid price $b_i$ for any sample $i$, and $\lambda \geq 0$ such that $\mathcal{L}$ is maximized and $\lambda \left( \sum_{i=1}^{N} F(x, b_i) C(b_i) - B \right) = 0$, then $b_i$ is the solution to Eq. (2) [18]. If $\lambda = 0$, $\mathcal{L}$ is maximized when $b_i \to +\infty$, and it is obviously impossible to submit an infinitely large bid price every time when bidding opportunities are sufficient and unlimited. Otherwise, the bidder will win all the opportunities and the cost will also be infinite contradicting the fact that the budget is finite. So $\lambda \left( \sum_{i=1}^{N} F(x, b_i) C(b_i) - B \right) = 0$ leads to the necessary conditions that $\lambda > 0$ and the budget is all spent as shown in Eq. (4):

$$\sum_{i=1}^{N} F(x, b_i) C(b_i) = B. \quad (4)$$

Secondly, we prove that estimating optimal control signal $\mu_0^*$ by campaign control and optimal bid price $b^*$ via bid shading in Eq. (3) are the necessary conditions of the optimization problem in Eq. (2).

According to the Karush-Kuhn-Tucker conditions [4], we know $\frac{\partial \mathcal{L}}{\partial \lambda} = -\left( \sum_{i=1}^{N} F(x, b_i) C(b_i) - B \right) = 0$. So the optimal lambda $\lambda^*$ is $\lambda^* = \arg\min_{\lambda_0} \left( \sum_{i=1}^{N} F(x, b_i) C(b_i) - B \right)$. We generate a control signal with lambda as $\mu_0 = 1/\lambda$. So the optimal control signal $\mu_0^*$ is:

$$\mu_0^* = \arg\min_{\mu_0} \left( \sum_{i=1}^{N} F(x, b_i) C(b_i) - B \right). \quad (5)$$

Since $F(x, b_i)$ and $C(b_i)$ are always positive, the cumulative cost $\sum_{i=1}^{N} F(x, b_i) C(b_i)$ is also positive. Consequently, the cumulative cost corresponding to the optimal control signal equals budget $B$.

Assume that bidders are independent of each other and we obtain the optimal lambda $\lambda^*$ (i.e., the optimal control signal $\mu_0^*$), we may cancel the sum in $\mathcal{L}$, ignore the constant $\lambda B$, and derive the optimal bid price $b^*$ for each sample independently as Eq. (6). Bid price is optimal when $\mathbb{E}(S) = (\mu_0^* V_i - C(b_i)) F(x, b_i)$ is maximum.

$$b_i^* = \arg\max_{b_i} (V_i - \lambda C(b_i)) F(x, b_i) = \arg\max_{b_i} \left( \frac{V_i}{\lambda_0^*} - C(b_i) \right) F(x, b_i)$$
$$= \arg\max_{b_i} (\mu_0^* V_i - C(b_i)) F(x, b_i). \quad (6)$$

Consequently, Eq. (5) and (6) are the necessary conditions for $b_i^*$ to be the optimal solution to Eq. (2).

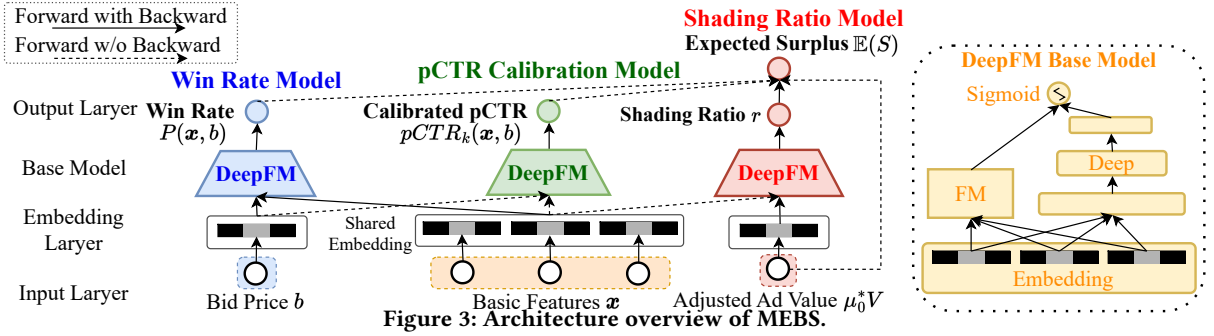

Figure 3: Architecture overview of MEBS.

Thirdly, we prove the uniqueness of the optimal bid price $b_i^*$ for any sample $i$ via proof by contradiction, and we conclude that the corresponding bidding strategy is equivalent to our aim in Eq. (2).

Let $(\mu_0^{*\prime}, b_i^{*\prime})$ also be optimal but different from $(\mu_0^*, b_i^*)$. We may assume that $\mu_0^{*\prime} > \mu_0^*$ without loss of generality, and there is a positive number $\delta$ satisfying $\mu_0^{*\prime} = \mu_0^* + \delta$. With different control signals $\mu_0^*$ and $\mu_0^{*\prime}$, the expected surpluses for any bid price $b_i$ are $\mathbb{E}(S_i) = (\mu_0^* V_i - C(b_i))F(\mathbf{x}, b_i)$ and $\mathbb{E}(S_i)' = (\mu_0^{*\prime} V_i - C(b_i))F(\mathbf{x}, b_i) = (\mu_0^* V_i - C(b_i))F(\mathbf{x}, b_i) + \delta V_i F(\mathbf{x}, b_i) = \mathbb{E}(S_i) + \delta V_i F(\mathbf{x}, b_i)$. The optimal bid prices $b_i^{*\prime}$ and $b_i^*$ are then defined in the following equations:

$$b_i^* = \arg\max_{b_i} \mathbb{E}(S_i) = \arg\max_{b_i} (\mu_0^* V_i - C(b_i))F(\mathbf{x}, b_i),$$

$$b_i^{*\prime} = \arg\max_{b_i} \mathbb{E}(S_i') = \arg\max_{b_i} (\mathbb{E}(S_i) + \delta V_i F(\mathbf{x}, b_i)).$$

As we assume the expected surplus $\mathbb{E}(S_i)$ is unimodal, $\mathbb{E}(S_i)$ monotonically increases when $b_i < b_i^*$ and monotonically decreases when $b_i > b_i^*$. Besides, $F(\mathbf{x}, b)$ shown in Figure 2 is strictly monotonically increasing, thus $\delta V_i F(\mathbf{x}, b_i)$ also monotonically increases. So $\mathbb{E}(S_i')$ monotonically increases when $b_i < b_i^*$, thus $b_i^{*\prime} \geq b_i^*$. If $b_i^{*\prime} = b_i^*$, $b_i^{*\prime}$ or $b_i^*$ is the unique optimal bid price. However, if $b_i^{*\prime} > b_i^*$, it implies that the cumulative expected cost also follows a strict inequality relationship $\sum_{i=1}^N C(b_i^{*\prime})F(\mathbf{x}, b_i^{*\prime}) > \sum_{i=1}^N C(b_i^*)F(\mathbf{x}, b_i^*)$ because $C(b_i^{*\prime}) \geq C(b_i^*)$ and $F(\mathbf{x}, b_i^{*\prime}) > F(\mathbf{x}, b_i^*)$. This contradicts Eq. (4) that the cumulative expected cost corresponding to the optimal control signal should equal the budget. Consequently, we finish the proof and prove $b_i^*$ is the unique optimal bid price.

In conclusion, we demonstrate the optimal bidding strategy for Eq. (2) is equivalent to optimizing campaign control and bid shading shown in Eq. (3). As for bid shading, the objective is to find the unique optimal bid price with the maximum expected surplus. □

## 5 Methodologies

In this section, we provide an overview of MEBS in Section 5.1 first. And we introduce each sub-module of MEBS. Finally, we illustrate the training and inference of MEBS in Section 5.5.

### 5.1 Overview of MEBS

Figure 3 depicts the architecture of MEBS. We construct three models: win rate model, pCTR calibration model, and shading ratio model. To alleviate the data sparsity of pCTR calibration and shading ratio estimation, we combine these models with multi-task learning and share embedding. With pre-trained win rate and pCTR calibration models, we calculate the expected surplus and maximize it in an end-to-end way optimizing shading ratio model implicitly.

### 5.2 Win Rate Model

To predict whether the auction will be won with bid price $b$, we build a win rate model to estimate the probability of winning $P(\mathbf{x}, b)$.

The win rate model is input with the bid price $b$ and basic features $\mathbf{x}$. Basic features include user features, impression features, ad features, and contextual features. The embedding layer gets the embeddings of these features, which are concatenated together and fed to the base model. The win rate model utilizes DeepFM [10] as a base model and estimates the win rate with a Sigmoid function.

We regard the win rate estimation as a binary classification task, which is supervised with the label of whether the auction is won $y_{\text{WR}}$. Cross-entropy loss is used for training and it is defined as:

$$L_{\text{WR}} = -\frac{1}{N}\sum_{i=1}^{N} y_{\text{WR}} P(\mathbf{x}, b) + (1 - y_{\text{WR}})(1 - P(\mathbf{x}, b)).$$

The win rate model is trained with both won and lost samples in the auction. Thus, the win rate model is less affected by data sparsity problem, and its feature embedding is well-trained.

### 5.3 Calibration of pCTR

Since the bid price influences the obtained ad position and CTR for MSDA, we conduct a bid-aware calibration of pCTR.

The pCTR for a given ad position as shown in Eq. (1) is affected by the impression itself and the ad position. The ad-specific and slot-independent $pCTR(\mathbf{x})$ is estimated by an upstream model in the impression valuation part before bidding, so it is not aware of the ad position determined by bid price. Given that the slot-specific factor in Eq. (1) is not differentiable, we model it in an end-to-end paradigm instead of using the average actual CTR. Our model inherently learns which ad slot to win and how it affects pCTR for any bid price, and it estimates a bid-aware calibration factor. We regard the logit of the base model as the calibration factor, and it is the sum of outputs from the FM layer and deep layer in a DeepFM model. First, assuming that the k-th slot is won, we may get the calibration factor $f_k(\mathbf{x}, b)$. Next, we get the original logit of $pCTR(\mathbf{x})$ from the inverse function of Sigmoid: $\log\left(\frac{pCTR(\mathbf{x})}{1-pCTR(\mathbf{x})}\right)$, and we add it with our calibration factor $f_k(\mathbf{x}, b)$. Finally, we feed the sum of the original logit and calibration factor to a Sigmoid function and estimate the pCTR of the k-th slot $pCTR_k(\mathbf{x}, b)$:

$$pCTR_k(\mathbf{x}, b) = \text{Sigmoid}\left(f_k(\mathbf{x}, b) + \log\left(\frac{pCTR(\mathbf{x})}{1 - pCTR(\mathbf{x})}\right)\right).$$

However, pCTR calibration faces data sparsity problem. Whereas the samples that win auctions and get impression opportunities are fairly rare, the samples clicked by users are far fewer. We alleviate the problem by sharing the embedding of win rate model with pCTR calibration model. Note that the shared embedding is frozen, and it only has forward propagation.

We train the pCTR calibration model with a cross-entropy loss. On winnable samples, we calculate the loss with label $y_{\text{Calib}}$ as:

$$L_{\text{Calib}} = -\frac{1}{N_+}\sum_{i=1}^{N_+} y_{\text{Calib}} pCTR_k(\mathbf{x}, b) + (1 - y_{\text{Calib}})(1 - pCTR_k(\mathbf{x}, b)).$$

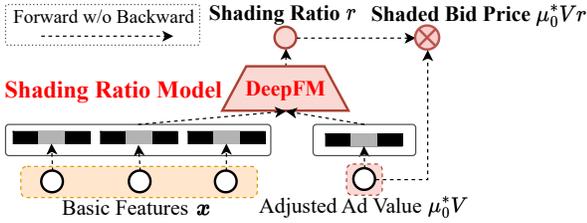

Figure 4: Inference of MEBS.

## 5.4 Shading Ratio Model

As shown in Figure 3, the shading ratio model estimates the optimal shading ratio and calculates the expected surplus with a DeepFM.

The inputs are adjusted ad value (i.e., the unshaded bid price) and basic features. Since shading ratio is only defined in winnable samples, data sparsity also hinders shading ratio model and is alleviated by sharing the embedding of win rate model.

The label of optimal shading ratio is unknowable in MSDA, and our model cannot be trained with the mean squared error of the estimated and the label of shading ratio like shading ratio regression method [9]. Our shading ratio model learns the distribution of the optimal bid price and it is implicitly optimized by maximizing the expected surplus in an end-to-end paradigm. With pre-trained win rate and pCTR calibration models, we could calculate the expected surplus as shown in Eq. (7) for every estimated shading ratio.

$$\mathbb{E}(S) = (\mu_0^* V - C(\mu_0^* \cdot V \cdot r)) P(x, \mu_0^* \cdot V \cdot r) pCTR_k(x, \mu_0^* \cdot V \cdot r). \quad (7)$$

The negative expected surplus loss is $L_{\text{surplus}} = -\frac{1}{N_+} \sum_{i=1}^{N_+} \mathbb{E}(S_i)$. And then the expected surplus is maximized with gradient descent. In first-price auctions, cost equals bid price $C(b) = b$, and the loss is of course differentiable. As for second-price auctions, we may substitute the cost $C(b)$ with the multiplication of bid price $b$ and cost-bid ratio $r_{\text{cb}}$ as $C(b) = b \cdot r_{\text{cb}}$. The cost-bid ratio is the statistical average of the according payment divided by the bid price, and it could be calculated from statistics in advance. Besides, we also apply weights and even the scale of loss $L_{\text{surplus}}$ in case it is dominated by samples with extremely large $\mathbb{E}(S)$ via dividing it with $\mathbb{E}(S)$ whose gradient is stopped. The final shading ratio model loss is defined as:

$$L_{\text{SR}} = -\frac{1}{N_+} \sum_{i=1}^{N_+} \frac{\mathbb{E}(S_i)}{\text{stop\_grad}(\mathbb{E}(S_i))}. \quad (8)$$

## 5.5 Training and Inference of MEBS

**5.5.1 Training of MEBS:** MEBS is trained by optimizing the above three models in order. First, we train win rate model on all samples in the auction with loss $L_{\text{WR}}$. Next, the pCTR calibration model is trained with loss $L_{\text{Calib}}$, and its feature embedding is shared with the win rate model. Finally, we maximize the expected surplus and optimize the shading ratio model implicitly with loss $L_{\text{SR}}$.

Following the idea of multi-task learning, MEBS comprises three closely intertwined tasks implicitly estimating the bidding landscape. Only when the win rate model estimates the minimum winning price well can it judge whether the bid price is able to win the auction. The pCTR calibration forecasts the relative position of bid price in the bidding environment and then finds the possible ad slot to win. Given that the optimal bid price is not too high or too low, the shading ratio model is also implicitly trained corresponding to the bidding environment. Since these tasks share a common goal, employing multi-task learning is not contradictory for them.

**5.5.2 Inference of MEBS:** The inference of MEBS shown in Figure 4 only requires shading ratio estimation. Unlike two-step bid shading methods searching for optimal bid price repeatedly, MEBS estimates shading ratio and directly gets bid price by multiplying the shading ratio with the adjusted ad value, ensuring significantly better inference efficiency.

## 6 Experiments

### 6.1 Offline Experiments

#### 6.1.1 Experiment Setup

**Dataset:** We utilize a bidding dataset sampled from the Alibaba display ad platform, and it consists of data from multi-slot second-price auctions in two days. The data on the first day and the next day serve as the training and test datasets respectively. And the statistics of the dataset are shown in Table 2. The samples win the auctions (#Won) or get clicked (#Clicked) are far less than the total samples (#Total) due to data sparsity. Samples in the dataset are converted from bid logs in our ad platform, and they include labels and features. Features contain bid price and other basic features, such as impression, ad, and contextual features. The labels consist of whether auctions are won and whether ads are clicked.

Table 2: Statistics of Dataset.

| #Total | #Won | #Clicked |
|---|---|---|
| 64557179576 | 533812872 | 9641171 |

**Implementation Details:** We set the embedding dim of sparse as 32. MEBS uses DeepFM as a base model, whose deep part is a three-layer MLP with 256, 64, and 16 nodes each. For model training, we employ Adam optimizer [15] and set the learning rate to 1e-2. We set batch size of win rate model to 80,000, and batch sizes of pCTR calibration and shading ratio model are 40,000.

**Evaluation Criteria:** Since we prove that the aim of bid shading is to maximize the expected surplus in Theorem 4.1, we adopt surplus in Eq. (9) as our metric for overall performance evaluation.

$$S = (\mu_0^* V - C(b)) \mathbb{I}(b \geq \text{wp}) pCTR_k(x, b), \quad (9)$$

where $C(b)$ is the bid price that wins the next ad position under the GSP setting on our platform, $\mathbb{I}(b \geq \text{wp})$ is an indicator function of whether the auction is won, and $pCTR_k(x, b)$ shown in Eq. (1) is the pCTR calibrated based on the actual ad positions before and after bid shading. When the slot-specific factor of $pCTR_k(x, b)$ is calculated based on the average CTR in corresponding ad **p**osition and **s**cene, we get Surplus (P&S). When the slot-specific factor is only calculated for each ad **p**osition, we get Surplus (P).

And PCOC (predicted CTR over the true CTR) for ablation study examines if calibrated pCTR aligns with actual CTR. The closer the PCOC is to 1, the better the performance of pCTR calibration.

**Baselines:** We implement bid shading baselines as introduced in Section 2.1, including shading ratio regression (SRR) method [9], two-step bid shading (TSBS) methods [23, 43] and non-parametric method (NPM) [39]. As for TSBS methods, we denote the study by Zhou et al. [43] as "TSBS-EDDN" and name the paper by Pan et al. [23] as "TSBS-WR". To mitigate the influence of the base model, we implement these methods with DeepFM employed by MEBS apart from original base models utilized by these baselines, such as LR[34], FwFM[22], and FM [24]. For non-censored methods relying on the optimal bid price label, we substitute the label with the bid price that wins the bottom ad slot (i.e., the minimum winning price).

#### 6.1.2 Overall Performance Comparison

The overall experiment results are shown in Table 3. MEBS shows at least a 9.7% lift on Surplus (P&S) and a 6.6% lift on Surplus (P), demonstrating superior performance compared with the baseline models, including non-censored methods and censored methods.

Table 3: Overall Offline Experiment Results.

| Category | Model | Surplus(P&S) | Surplus(P) |
|---|---|---|---|
| Our Method | MEBS | **16.020743** | **11.634698** |
| Non-censored | SRR(DeepFM) | 14.467197 | 10.914265 |
|  | SRR(FM) | 13.858914 | 9.267927 |
|  | TSBS-EDDN(DeepFM) | 9.339563 | 6.174544 |
|  | TSBS-EDDN(FwFM) | 9.797970 | 5.743973 |
|  | NPM | 10.234589 | 6.637969 |
| Censored | TSBS-WR(DeepFM) | 10.339686 | 6.454219 |
|  | TSBS-WR(LR) | 8.089951 | 4.865804 |
|  | TSBS-EDDN(DeepFM) | 9.826253 | 6.131234 |
|  | TSBS-EDDN(FwFM) | 9.288596 | 5.743973 |

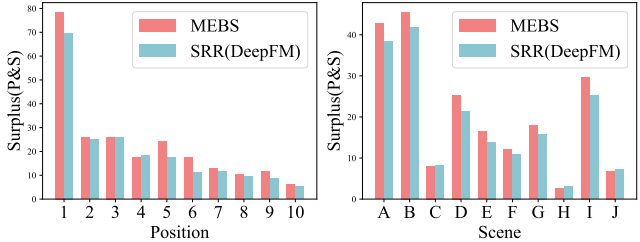

Figure 5: Surplus in Top Ad Positions and Main scenes.

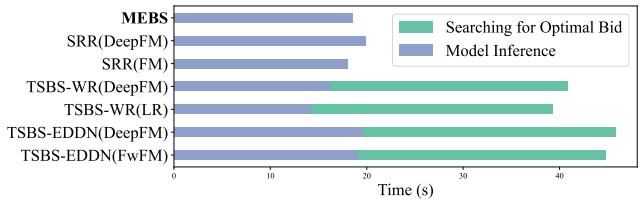

Figure 6: Inference Efficiency Comparison.

As shown in Figure 5, we analysis overall surplus performance in the top 10 ad positions and the 10 scenes with the highest revenue in Taobao App, such as Guess What You Like and After Purchase, etc. Compared with the best baseline SRR(DeepFM) in Table 3, MEBS performs better in almost every top positions, especially in the first ad slot. And MEBS also shows superiority in most listed scenes.

We also compare the inference efficiency of MEBS with other parametric baselines in the same computing environment, and the results are shown in Figure 6. After saving computing graphs and reloading from checkpoints, we only count operator execution time, eliminating I/O and metric calculation interference. We set the batch size to 5,000,000 and calculate the average inference time per batch in seconds. Without the searching process, MEBS is more efficient than TSBS due to end-to-end learning. MEBS and SRR both only estimate shading ratio, resulting in zero searching time.

In conclusion, MEBS achieves superior performance while maintaining excellent efficiency.

### 6.1.3 Ablation Study

The results of our ablation study are shown in Table 4. After removing embedding sharing, we get the results of MEBS without multi-task learning, whose surpluses decline 15.5~21.1%. As for MEBS without end-to-end learning, we substitute its shading ratio estimation loss with mean squared error applied by the SRR baseline, and its surpluses degenerate to the level of SRR. To examine the effect of pCTR calibration, we also make an ablation by removing it from MEBS, and we observe drops in surplus. Additionally, after removing pCTR calibration, the PCOC of MEBS degenerates from 1.034566 to 1.104969, indicating a degradation of pCTR accuracy. The results demonstrate the effectiveness of pCTR calibration.

Table 4: Ablation Study Results.

| Model | Surplus(P&S) | Surplus(P) |
|---|---|---|
| MEBS | **16.020743** | **11.634698** |
| MEBS without Multi-task Learning | 13.544584 | 9.174031 |
| MEBS without End-to-end Learning | 14.757131 | 10.623542 |
| MEBS without pCTR Calibration | 14.536990 | 10.051408 |

Table 5: Online Experiment Results.

| Metric | GMV | ROI | BuyCnt | CPC |
|---|---|---|---|---|
| A/B result | +7.01% | +7.42% | +3.26% | -8.37% |

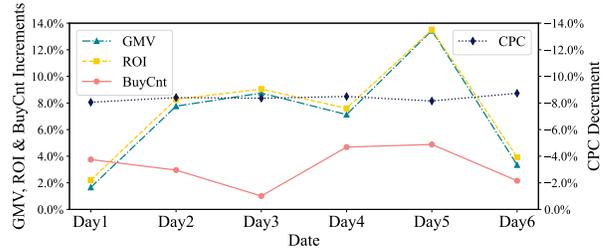

Figure 7: Online Performance of Each Day.

### 6.2 Online Experiments

In online A/B tests, we use gross merchandise volume (GMV), return on investment (ROI), buy count (BuyCnt), and cost per click (CPC) as our metrics. GMV indicates the total gained revenue. ROI is calculated by dividing GMV by the cost showing the cost-effectiveness of advertising. Buy count is the total number of ads that are brought. CPC is the average cost for a click, implying the payment level of bidding. As for CPC, the lower the better.

We experiment on the Alibaba display ad platform for six days. In A/B tests, we choose 10,000 ad campaigns and spare 25% ad requests for experiments. These ad campaigns and requests are equally divided into two groups. One group receives the control and the other group receives the treatment of MEBS. The percentage change results of metrics are shown in Table 5. The lift results of GMV, ROI, and BuyCnt as well as the decline of CPC prove that MEBS improves advertising performance.

From the results of each day shown in Figure 7, we demonstrate that MEBS performs well during the experiment period. The increment percentages of GMV, ROI, and BuyCnt stay positive each day. And the change percentage of CPC stabilizes around -8%, indicating MEBS saves costs for advertisers steadily.

## 7 Conclusion

We propose the MEBS model, which is the first to adapt the bid shading method to multi-slot display advertising. We prove the optimality of our bidding strategy with bid shading. Leveraging multi-task learning, we mitigate the issue of data sparsity that hampers pCTR calibration and shading ratio estimation. Following the objective of bid shading in our proof, we directly maximize the expected surplus to optimize shading ratio estimation in an end-to-end paradigm. Through extensive offline and online experiments, we prove the effectiveness and efficiency of MEBS and show its significant impact on enhancing bidding performance.